\documentclass[aps,prl,twocolumn,showpacs,groupedaddress]{revtex4}  % for review and submission 
\usepackage{graphicx}  % needed for figures
\usepackage{dcolumn}   % needed for some tables
\usepackage{bm}        % for math
\usepackage{amssymb}   % for math
\usepackage{axodraw}   % for feyn pictures

\newcommand{\bbb}{$b\bar{b}$}

\newcommand{\invfb}{fb$^{-1}$}

\newcommand{\met}{\mbox{\ensuremath{\,\slash\kern-.7emE_{T}}}}

\newcommand{\mht}{\mbox{\ensuremath{\,\slash\kern-.7emH_{T}}}}

\def\lsim{\mathrel{\rlap{\lower4pt\hbox{$\sim$}}
    \raise1pt\hbox{$<$}}}                % less than or approx. symbol

\newcommand{\sq}{\tilde{q}}

\newcommand{\st}{\tilde{t}}

\def\dzero{D0}
\def\m0{$m_0$}
\def\mf{$m_{1/2}$}
\def\sq{$\tilde q$}
\def\gl{$\tilde g$}
\def\neuu{$\tilde \chi _1 ^0$}
\def\neud{$\tilde \chi _2 ^0$}
\def\chau{$\tilde \chi _1 ^\pm$}

\def\st1{$\tilde \tau _1 ^\pm$}
\def\stu{$\tilde \tau _1 ^\pm$}

\def\htt{$H_T$}

\begin{document}
%\setpagewiselinenumbers
%\modulolinenumbers[5]
%\linenumbers

% the following line is for submission 
\hspace{5.2in} \mbox{FERMILAB-PUB-09/268-E}

\title{
Search for squark production in events with jets, hadronically decaying tau leptons
and missing transverse energy
at $\bm{\sqrt{s}}$=1.96 TeV
}
% LIST_OF_AUTHORS_R2.TEX                 5/15/09            
%
\author{V.M.~Abazov$^{37}$}
\author{B.~Abbott$^{75}$}
\author{M.~Abolins$^{65}$}
\author{B.S.~Acharya$^{30}$}
\author{M.~Adams$^{51}$}
\author{T.~Adams$^{49}$}
\author{E.~Aguilo$^{6}$}
\author{M.~Ahsan$^{59}$}
\author{G.D.~Alexeev$^{37}$}
\author{G.~Alkhazov$^{41}$}
% ...

\affiliation{\vspace{0.1 in}(The D\O\ Collaboration)\vspace{0.1 in}}
% ...
\affiliation{$^{6}$University of Alberta, Edmonton, Alberta, Canada;
                Simon Fraser University, Burnaby, British Columbia, Canada;
                York University, Toronto, Ontario, Canada and
                McGill University, Montreal, Quebec, Canada}
% ...
\affiliation{$^{30}$Tata Institute of Fundamental Research, Mumbai, India}
% ...
\affiliation{$^{37}$Joint Institute for Nuclear Research, Dubna, Russia}
% ...
\affiliation{$^{41}$Petersburg Nuclear Physics Institute,
                St. Petersburg, Russia}
% ...
\affiliation{$^{49}$Florida State University, Tallahassee, Florida 32306, USA}
% ...
\affiliation{$^{51}$University of Illinois at Chicago,
                Chicago, Illinois 60607, USA}
% ...
\affiliation{$^{59}$Kansas State University, Manhattan, Kansas 66506, USA}
% ...
\affiliation{$^{65}$Michigan State University,
                East Lansing, Michigan 48824, USA}
% ...
\affiliation{$^{75}$University of Oklahoma, Norman, Oklahoma 73019, USA}
% ...
  
\date{May 25, 2009}
%\date{\today}

\begin{abstract}
A search for supersymmetric partners of quarks is performed in the topology of multijet events accompanied by at least
one tau lepton decaying hadronically and large missing transverse energy. Approximately 
1\,fb$\rm ^{-1}$ of $p\bar{p}$ collision data from the Fermilab Tevatron Collider at a
center of mass energy of 1.96 TeV recorded by the \dzero\ detector is analyzed.
Results are combined with the previously published \dzero\ inclusive search for squarks and gluinos.
No evidence of physics beyond the standard model is found and lower limits on the squark mass
up to 410~GeV are derived in the framework of minimal supergravity with $\tan\beta=15$, $A_0=-2m_0$ and $\mu<0$,
in the region where decays to tau leptons dominate. Gaugino masses $m_{1/2}$ are excluded up to 172~GeV. 
\end{abstract}

\maketitle 

%%%%%%%%%%%%%%%%%%%%%%%%%%%%%%%%%%%%%
% Introduction
%%%%%%%%%%%%%%%%%%%%%%%%%%%%%%%%%%%%%

Supersymmetric (SUSY) extensions \cite{susytheo} of the standard model (SM) predict the existence 
of scalar quarks (\sq), or squarks, and fermionic gluons (\gl), or gluinos, the supersymmetric partners 
of quarks and gluons. These particles carry color and, if they are sufficiently light, they could be the 
most copiously produced SUSY particles at hadron colliders.
The mass reach in direct searches at the CERN $e^+e^-$ Collider (LEP) is generally expected to be lower.
However, direct searches for charginos and sleptons at LEP place stringent bounds in SUSY parameter space 
which do translate into indirect constraints on squarks and gluinos \cite{susyreview}.
At the Fermilab Tevatron Collider, previous searches for squarks and gluinos assuming R-parity 
conservation~\cite{cdf:sqgl,verdier:2008} were performed using events with jets accompanied by large missing
transverse energy arising from the undetected lightest supersymmetric particle (LSP) assumed to be the lightest 
neutralino ({\neuu}). However, in some regions of the parameter space, squark and gluino cascade decays lead 
to final states with leptons. 

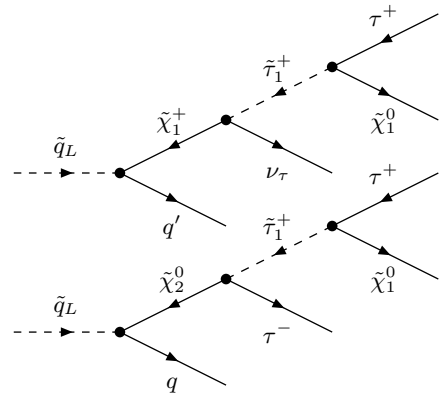
\begin{figure}[htb]
\begin{center}
\begin{picture}(220,160)(0,0)
%%%%%%%%%%%%%%%%%%%%%%%%%%%%%%%%%%%%%%%%%%%%%%%%%%%%%%%%%%%%%%%%%%%%%%%%%%%%%%%%%
% cadre
%\Line(0,0)(0,160)
%\Line(0,160)(220,160)
%\Line(220,160)(220,0)
%\Line(220,0)(0,0)
%%%%%%%%%%%%%%%%%%%%%%%%%%%%%%%%%%%%%%%%%%%%%%%%%%%%%%%%%%%%%%%%%%%%%%%%%%%%%%%%%
% squark L
\DashArrowLine(20,30)(60,30){3}{}
\Text(40,40)[]{$\tilde{q}_{L}$}
% chargino1 - q'
\ArrowLine(100,50)(60,30)
\Text(80,50)[]{$\tilde{\chi}^{0}_{2}$}
\ArrowLine(60,30)(100,10)
\Text(80,10)[]{$q$}
% stau1 - nu tau
\DashArrowLine(140,70)(100,50){3}{}
\Text(120,70)[]{$\tilde{\tau}^{+}_{1}$}
\ArrowLine(100,50)(140,30)
\Text(120,30)[]{$\tau^{-}$}
% tau+ - chi10
\ArrowLine(180,90)(140,70)
\Text(160,90)[]{$\tau^{+}$}
\ArrowLine(140,70)(180,50)
\Text(160,50)[]{$\tilde{\chi}^{0}_{1}$}
\Vertex(60,30){2}
\Vertex(100,50){2}
\Vertex(140,70){2}
%
%%%%%%%%%%%%%%%%%%%%%%%%%%%%%%%%%%%%%%%%%%%%%%%%%%%%%%%%%%%%%%%%%%%%%%%%%%%%%%%%%
% squark L
\DashArrowLine(20,90)(60,90){3}{}
\Text(40,100)[]{$\tilde{q}_{L}$}
% chargino1 - q'
\ArrowLine(100,110)(60,90)
\Text(80,110)[]{$\tilde{\chi}^{+}_{1}$}
\ArrowLine(60,90)(100,70)
\Text(80,70)[]{$q'$}
% stau1 - nu tau
\DashArrowLine(140,130)(100,110){3}{}
\Text(120,130)[]{$\tilde{\tau}^{+}_{1}$}
\ArrowLine(100,110)(140,90)
\Text(120,90)[]{$\nu_{\tau}$}
% tau+ - chi10
\ArrowLine(180,150)(140,130)
\Text(160,150)[]{$\tau^{+}$}
\ArrowLine(140,130)(180,110)
\Text(160,110)[]{$\tilde{\chi}^{0}_{1}$}
\Vertex(60,90){2}
\Vertex(100,110){2}
\Vertex(140,130){2}
\end{picture}
\caption{Main Feynman graphs of left handed squark cascade decays into tau leptons; one tau lepton or two tau leptons are produced if squarks decay through a \chau\ or a \neud.}
\label{fig:decayscheme}
\end{center}
\end{figure}

In the case of the third generation of sfermions, large mixing between the left and right handed chiral states may occur.
Thus, in a given model, the lighter mass state of supersymmetric partners of the tau lepton (\stu) might be 
the lightest of all sleptons \cite{susyreview}. In addition, if the \stu\ is lighter than charginos (\chau) 
and neutralinos (\neud), they decay exclusively into staus, $ \tilde \chi _1 ^\pm \to \tilde \tau_1 ^\pm \nu_{\tau}$ 
and $\tilde \chi_2^0 \to \tilde \tau_1 ^\pm \tau^\mp$, with the subsequent decay $\tilde \tau_1 ^\pm \to \tau^\pm \tilde \chi _1 ^0$,
leading to final states with tau leptons. Other decays to electrons and muons would be suppressed. 
In this region, sometimes refered to as the ``tau corridor'', squarks are lighter than gluinos and the
production of squark pairs of the two first generations, $p\bar{p} \to \tilde q_R \tilde q_L$, dominates.
Right handed squarks decay into a quark and the LSP while left handed squarks decay into a quark accompanied 
by the lightest chargino about 2/3 of the time or the second lightest neutralino about 1/3 of the time.
Figure~\ref{fig:decayscheme} illustrates typical decay chains into final states with tau leptons. 
These cascade decays are characterized by the presence of two or more jets, large missing transverse energy 
(\met) and at least one tau lepton. Such signatures have not previously been exploited in SUSY searches at 
the Fermilab Tevatron Collider.

In this Letter, we report on a search for final states containing at least two jets, at least one tau lepton decaying hadronically and large missing transverse energy using 0.96~\invfb\ of data collected in $p \bar p$ collisions at a center of mass energy of $\sqrt{s}=1.96$ TeV with the upgraded \dzero\ detector during Run~II of the Fermilab Tevatron Collider.
Squark production is investigated in the framework of minimal supergravity (mSUGRA) \cite{msugra} with parameters enhancing final states with tau leptons.
This analysis is further combined with the previously published D0 analyses \cite{verdier:2008} developed in the framework of inclusive searches for squark and gluino production.

%%%%%%%%%%%%%%%%%%%%%%%%%%%%%%%%%%%%%
% Detector
%%%%%%%%%%%%%%%%%%%%%%%%%%%%%%%%%%%%%

The \dzero\ detector~\cite{nimd0detector} consists of a central tracking system, inside a calorimeter and a muon detector.
The tracking apparatus is installed within a superconducting solenoidal magnet of 2~T. It is composed of two subsystems, a silicon microstrip tracker and a central fiber tracker, which provide efficient tracking and vertexing over the pseudorapidity range $|\eta_{}| \le $ 2.5, where $\eta_{} = - {\rm ln[tan}~ (\theta /2)]$,  $\theta$ being the polar angle with respect to the proton beam direction $z$. 
The calorimeters use a sampling technology with liquid argon and depleted uranium organized in projective towers of size $0.1 \times 0.1$ in the $\eta - \phi$ plane, where $\phi$ is the azimuthal angle. 
The central calorimeter ($|\eta| < 1.1$) is separated from the end calorimeters which extend the pseudorapidity coverage up to $\sim$ 4.2. Each calorimeter is located in its own cryostat, creating gaps in the interval $1.2 < |\eta| < 1.4$, resulting in a degradation in the energy resolution for jets in this region.
The inter cryostat detector (ICD) is instrumented with scintillating tiles. It provides additional coverage in the range $1.1 < |\eta| < 1.4$.
Calorimeter towers are subdivided into fine longitudinal layers. The ones closest to the interaction point form the electromagnetic (EM) section of the calorimeters, followed by the hadronic layers which include the ICD.
The muon detector surrounds the calorimeter. It provides coverage up to  $|\eta_{}| \approx $ 2.0. It is equipped with three layers of tracking detectors and scintillating trigger counters.
An iron toroidal magnet of 1.8~T is installed around the innermost layer.

The data analyzed in this search were collected from April 2003 through February 2006. 
Events were selected online with unprescaled calorimeter based triggers designed to select dijet and multijet events with missing transverse energy. 
At trigger level, there are no requirements on calorimeter clusters that would reject hadronic tau lepton decays, so these clusters will include both jets and tau candidates.
The data sample corresponds to an integrated luminosity of $(0.96 \pm 0.06)$~\invfb~\cite{andeen}.
Due to the increasing instantaneous luminosity of the Fermilab Tevatron Collider, trigger requirements are tightened with time. In this Letter, we describe the most recent trigger requirements used. 
The \dzero\ triggering system is organized in three successive levels (L1, L2 and L3). Events are selected by the L1 if there are at least three trigger towers (of size $0.2 \times 0.2$ in $\eta - \phi$ plane) with transverse energy greater than 5~GeV.
For dijet events, the modulus of the vector sum of the transverse momentum of simple cone jets, \mht, is required to be greater than 20~GeV and 30~GeV at L2 and L3, respectively.
The azimuthal angle between the two calorimeter clusters of highest transverse energy must be less than 168.75$^\circ$ and 170$^\circ$ at L2 and L3, respectively. In addition, at L3, the scalar sum of the transverse momentum of the jets, \htt, must exceed 50~GeV and the minimal distance in azimuth between any calorimeter cluster and \mht\ is required to be greater than 25$^\circ$.
For multijet events, three calorimeter clusters with transverse energy greater than 6~GeV and 20~GeV are required, and \htt\ must exceed 75~GeV and 125~GeV, at L2 and L3, respectively. At L3, \mht\ is required to be above 25~GeV.
The trigger efficiency is greater than 94\% for the signal events considered in this Letter that satisfy the offline selection criteria.

We consider the signal event signature with two or more jets and at least one hadronically decaying tau lepton, accompanied by large missing transverse energy. SM processes involving $W$ bosons decaying to a tau lepton and a neutrino, in association with jets, including top quark pair production and decay, contribute to the irreducible background. Another source of background is events with spurious \met\ and electrons, muons or jets mimicking taus. SM expectations and signal efficiencies are computed with Monte Carlo methods, except for the multijet background from QCD processes, which is evaluated directly from data.
The detector geometry and response are simulated with a program based on {\sc geant}~3 \cite{geant}.
One random beam crossing data event, recorded during data taking, is overlaid on each Monte Carlo event according to the instantaneous luminosity of the data sample in order to simulate detector noise and additional soft interactions. 

The hard scatter in SM processes is generated with version 2.05 of {\sc alpgen}~\cite{alpgen}, except for diboson and single top quark processes generated with {\sc pythia}~6.323~\cite{pythia} and {\sc comphep}~4.1.10~\cite{comphep}.
The parton distribution functions (PDF) are modeled using the {\sc{cteq6l1}}~\cite{cteq1,cteq2} library. 
In all cases, the initial and final state radiation and the parton hadronization are simulated by {\sc pythia}. 
For all {\sc alpgen} samples, the MLM parton-jet matching prescription~\cite{Hoche:2006ph} is applied to avoid duplicate phase space regions where both the matrix element as well as the parton shower contribute to the formation of parton jets.   
The inclusive tau lepton decay is simulated with the program {\sc tauola}~2.5~\cite{Jadach:1990mz} with slighltly modified tau lepton branching ratios motivated by \cite{pdg2004}.
Next-to-leading order (NLO) QCD corrections for SM processes are taken into account by applying {\sc mcfm}~5.1~\cite{mcfm} NLO correction factors to the leading order (LO) cross sections calculated by the event generators.

The analysis optimisation is performed in the mSUGRA model in a region of parameter space where final states with tau leptons dominate. While the universal scalar and gaugino masses, $m_0$ and $m_{1/2}$, are varied to explore the region of interest,
a large mixing in the stau mass matrix is obtained by fixing the ratio of the neutral Higgs vacuum expectation values, $\tan \beta$, to 15. The universal trilinear coupling, $A_0$, is set to $-2m_0$ to favor high Higgs boson masses and the sign of the Higgs mixing mass parameter $\mu$ is negative so that the lightest chargino mass is increased while squark pair production cross sections are unchanged. 
This helps evading the indirect limits set by LEP.
Squark and gluino pairs are produced, including all species but stop quarks, and their decay is simulated using {\sc pythia}. Sparticle masses and couplings are computed with {\sc suspect~2.3}~\cite{suspect23}, and {\sc sdecay}~1.1a~\cite{sdecay11a}, respectively. Samples are normalized to the NLO cross section computed with {\sc prospino}~2 \cite{prospino2} and the {\sc cteq6.1m} PDF set.
The signal cross section is typically $\sim$0.3 pb for squark masses of about 350~GeV.

Collision data and simulated events are processed through the same reconstruction chain.
Calorimeter towers are clustered in jets with a cone algorithm of radius ${\cal R}=\sqrt{(\Delta y)^2 + (\Delta \phi)^2 }=0.5$ \cite{runIIconealgo}, where $y= \frac{1}{2}{\rm ln} [(E+ p_z)/(E- p_z)]$ is the rapidity. Only jets with transverse momentum ($p_T$) above 13~GeV are considered. A jet energy scale correction (JES) is applied and simulated events are corrected to account for the measured jet reconstruction efficiency in data events.
Hadronic tau lepton decays are characterized by a narrow isolated calorimeter cluster with low track multiplicity \cite{conf:galea,zttdzero}. Tau finding starts with calorimeter clusters with $E_T \ge 5$~GeV constructed with a simple
cone algorithm with ${\cal R}=0.5$.
The innermost part of the cone (${\cal R}=0.3$) is used to measure the tau candidate transverse energy while the energy deposited between ${\cal R}=0.3$ and ${\cal R}=0.5$ is used to compute the tau isolation variable. 
Tracks of $p_T \ge$~1.5~GeV pointing to tau candidates in ${\cal R}=0.5$ are associated to them.
In addition, we also use isolated tracks with $p_T>5$~GeV as seeds for taus. In this case, a calorimeter cluster of ${\cal R}=0.5$ is constructed around the track and the rest of the tau construction procedure is the same as for the calorimeter seeded taus.
All tau candidates are required to be associated with at least one track. One or two additional tracks, within 2~cm in $z$ of the highest $p_T$ track at closest approach, are added if the track invariant mass does not exceed 1.1~GeV and 1.7~GeV respectively. 
Moreover, the third track is added only if the sum of the electric charge of the three tracks is equal to $\pm 1$. 
Electromagnetic sub-clusters with a minimum $E_T$ of 0.8~GeV are constructed from EM calorimeter cells belonging to the tau cluster.
Tau candidates are further classified into three types, based on the tau signature in the detector. Type-1 taus have one associated track and no EM sub-clusters, typical of the decay into an isolated single pion or kaon. Type-2 taus have one associated track and at least one EM sub-cluster. Type-2 taus come mainly from decays via $\rho^\pm$ or $a_1^\pm$ into a charged pion and one or more $\pi^0$'s. These decay channels represent more than 50\% of hadronic tau lepton decays.
They also include $\tau\to e$ decays.
Type-3 taus have more than one associated track, resulting from multiprong tau lepton decays.
The tau candidate transverse energy is inferred from the transverse momenta of the associated tracks, the $E_T$ of the calorimeter cluster and the known calorimeter response to $\pi^\pm$ parametrized as a function of track $p_T$ and $\eta_{}$.
The $\met$ is computed from calorimeter cells prior to clustering. It is corrected for the JES and the $p_T$ of reconstructed muons.

Preselected events are required to have at least two reconstructed jets and \met\ greater than 40~GeV.
The event primary interaction vertex (PV) is restricted to be within 60 cm of the detector center along the $z$ axis, $|z_{\rm PV}| < 60$~cm.
The two jets with highest $p_T$, $j_1$ and $j_2$, are required to be reconstructed in the central part of the calorimeter ($|\eta_{}| < 0.8$), to have $p_T \ge 35$~GeV and not to be collinear in azimuthal plane. Jet $p_T$ can be mismeasured if the primary vertex is incorrect, which can lead to apparent high \met.
This background from misreconstruction, as well as instrumental backgrounds from noise in the calorimeter, is significantly reduced by requiring, for both $j_1$ and $j_2$, $CPF0 \ge 0.75$, where $CPF0$ is the fraction of the sum of $p_T$ of charged particles in a jet that point to the PV compared to the sum of $p_T$ of charged particles in the
jet that point to any vertex.

Events selected by the dijet trigger or the multijet trigger have different topologies. This fact is exploited by performing two different analyses, referred to as ``tau-dijet'' and ``tau-multijet'', as summarized in Table~\ref{tab:taucutflow}.
In both selections, the \met\ requirement is tightened to 75~GeV.
In the ``tau-multijet'' analysis, a third energetic jet is required and events with $H_T$ less than 200~GeV are discarded.
Multijet events with one mismeasured jet leading to \met\ in the jet direction are rejected by selecting events with \met\ well separated in azimuth from the two leading jets.
In the ``tau-dijet'' analysis, this procedure is extended to any jet reconstructed in the event. 
Major background contributions are $W$ boson and top quark pair production, $Z(\to \nu\nu)+$jets events and multijet events.

\begin{table}[htbp]
\renewcommand{\arraystretch}{1.2}
\caption{Selection criteria for the ``tau-dijet'' and ``tau-multijet'' analyses.
\label{tab:taucutflow}}
\begin{ruledtabular}
\begin{tabular}{lcc}
Preselection & \multicolumn{2}{c}{All analyses}\\
\hline
\met         & \multicolumn{2}{c}{$\ge$ 40 GeV}\\
number of jets &\multicolumn{2}{c}{$\ge$ 2}\\
$\Delta \Phi (j_1 , j_2 )$ & \multicolumn{2}{c}{$<$ 165$^\circ$}\\
$|z_{\rm PV}|$     & \multicolumn{2}{c}{$<$ 60 cm}\\
$j_1$ $p_T$, $|\eta_{}|$, $CPF0$ &  \multicolumn{2}{c}{$\ge$ 35 GeV, $\le$ 0.8, $\ge 0.75$}\\
$j_2$ $p_T$, $|\eta_{}|$, $CPF0$ &  \multicolumn{2}{c}{$\ge$ 35 GeV, $\le$ 0.8, $\ge 0.75$}\\
\hline
Jets and \met\ selection & ``tau-dijet'' & ``tau-multijet'' \\
\hline
trigger & dijet & multijet \\
\met         &$\ge$ 75 GeV &$\ge$ 75 GeV\\
$\Delta \Phi ($\met$, {j_1})$&$ \ge $ 90$^\circ$ &$ \ge $ 90$^\circ$\\
$\Delta \Phi ($\met$, {j_2})$&$ \ge $ 50$^\circ$ &$ \ge $ 50$^\circ$\\
$\Delta \Phi_{\rm min} ($\met$, {\rm any~jet})$&$ \ge $ 40$^\circ$&-\\
number of jets & - & $\ge$ 3\\
$j_3$ $p_T$, $|\eta|$&-&  $\ge$ 35 GeV, $\le$ 2.5 \\
$H_T$ &-& $\ge$ 200 GeV\\
\hline
Tau candidate selection & ``tau-dijet'' & ``tau-multijet'' \\
\hline
number of tau candidates & {$\ge$ 1} & {$\ge$ 1}\\
$\Delta {\cal R} (\tau_{\rm cand}, {j_1})$ & {$\ge$ 0.5} & {$\ge$ 0.5}\\
$\Delta {\cal R} (\tau_{\rm cand}, {j_2})$ & {$\ge$ 0.5} & {$\ge$ 0.5}\\
\hline
Optimisation & \multicolumn{2}{c}{``tau-dijet'' \texttt{OR} ``tau-multijet''}\\
\hline
$S_T$  & \multicolumn{2}{c}{$\ge 325$~GeV}\\
\met  & \multicolumn{2}{c}{$\ge 175$~GeV}\\
\end{tabular}
\end{ruledtabular}
\end{table}

Events are required to contain at least one tau candidate not already identified as one of the two highest $p_T$ jets, $j_1$ and $j_2$. Only tau candidates with $E_T$ above 15~GeV are considered. The transverse momentum of the associated track in type-1 taus must exceed 4~GeV, and, in case of type-3 taus, the scalar sum of the $p_T$ of the  associated tracks is required to be greater than 8~GeV.
Quark or gluon jets are reconstructed as tau candidates with a probability between 0.2 and 0.8, strongly depending on energy and position in the detector. They are efficiently separated from hadronic tau lepton decays using Neural Networks ($NN_j$), one for each tau type, which exploit the difference in transverse and longitudinal shower shape, as well as the isolation in the tracker and in the calorimeter. Training is performed with simulated $Z\to \tau \tau$ as signal and tau candidates in a multijet enriched sample from data as background. 
This analysis uses comparatively efficient $NN_{j}$ cut values \cite{zttdzero} to ensure a high selection efficiency for hadronic tau lepton decays (75\%) while keeping a good background rejection (factor 14).
The choice of moderate $NN_j$ requirements is in response to the very high rejection power of $S_T$ and \met\ selections against SM backgrounds, where $S_T = p_T^{j_1} + p_T^{j_2} + E_T^{\tau}$. 
Electrons are generally reconstructed as type-2 taus. Another Neural Network was developed to separate electrons from hadronic tau lepton decays. In this case, $Z\to ee$ events are used for background training. The cut applied gives a rejection factor on electrons of about about 20 while keeping 95\% of hadronic tau lepton decays.
Electrons in the ICD are reconstructed as type-1 taus. Therefore, type-1 taus in this region are discarded.
Muons are detected as tracks in the central tracker and they leave a MIP-like signature in the calorimeter, which can be misidentified as type-1 taus. This contribution, as well as poorly reconstructed tracks, are suppressed for type-1 and type-3 taus by comparing the energy of the calorimeter cluster and the momentum of the associated track(s) ($E/p \ge 0.7$). In addition, the extremely high density of the {\dzero} calorimeter favors the interaction of muons by Bremsstrahlung in the outermost layers. We reject tau candidates if they deposit more than 40\% of their energy in those layers. This criterion also suppresses beam halo background and pions interacting by charge exchange in the calorimeter.

\begin{figure}[htbp]
\begin{center}
\includegraphics[scale=0.32,angle=00]{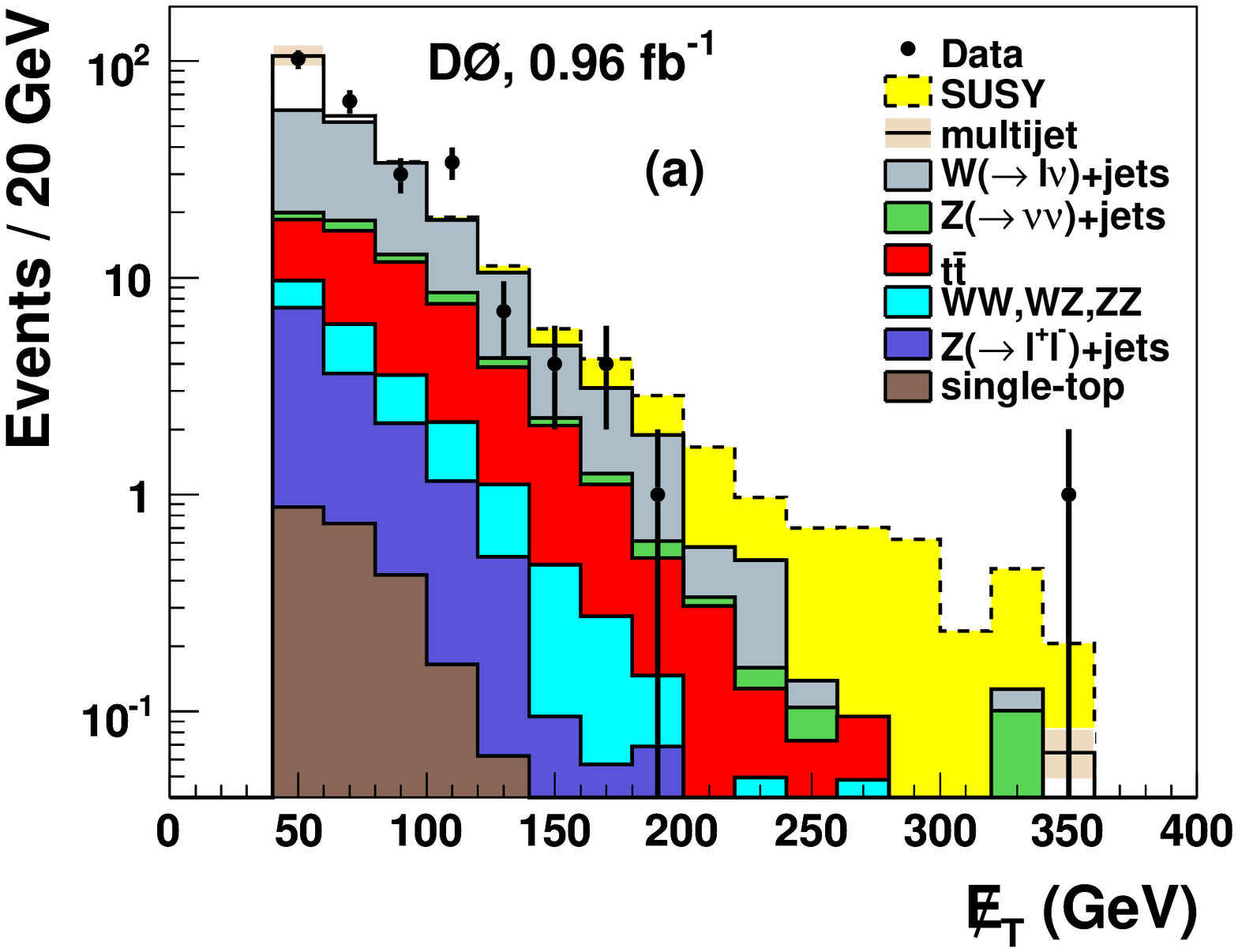}
\hfill
\includegraphics[scale=0.32,angle=00]{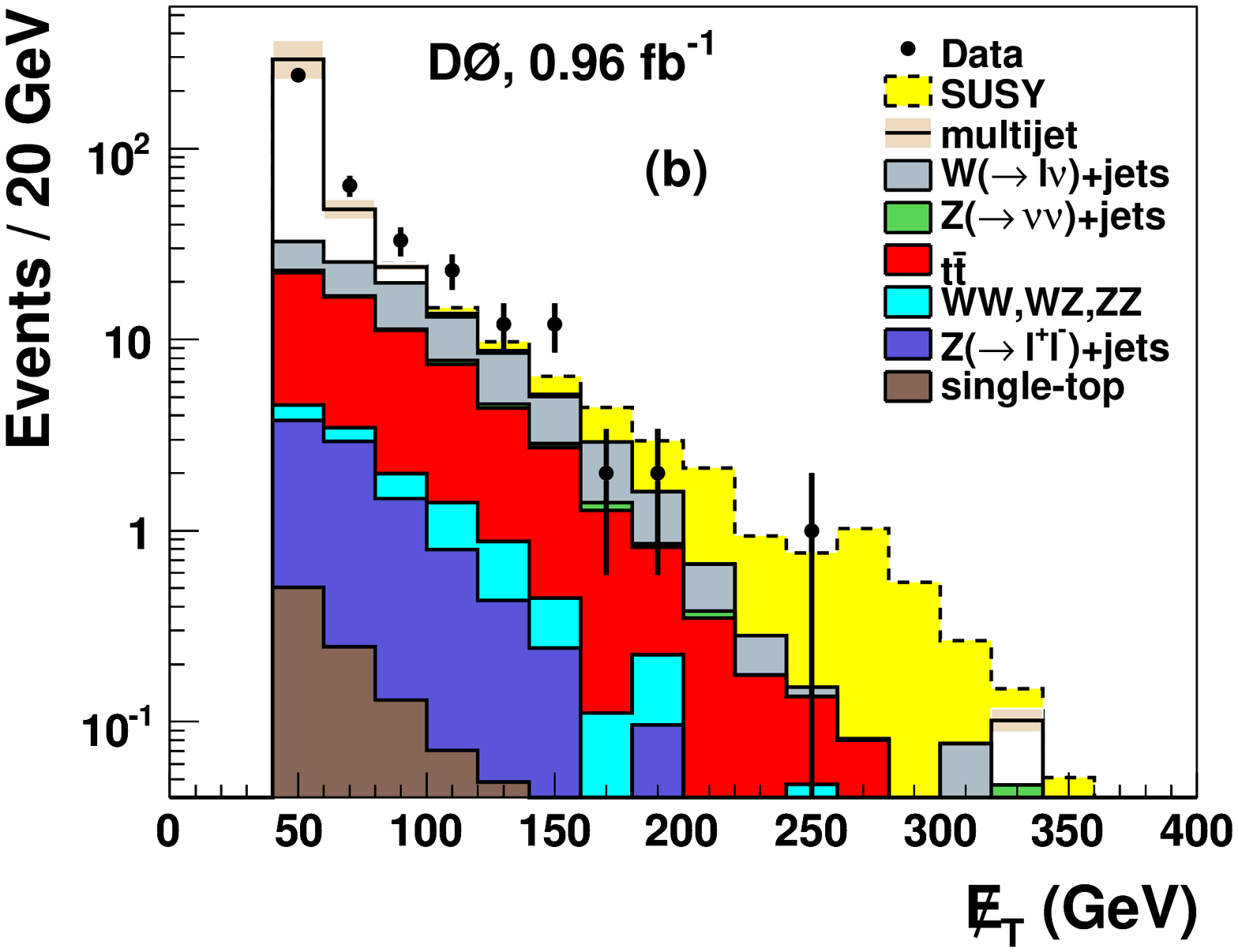}
\caption{Distributions of \met\ after the tau selection separately for the ``tau-dijet'' (a) and ``tau-multijet'' (b) analyses. The \met\ cut is relaxed to 40~GeV. Data (points with error bars), SM predictions (plain histograms) with the estimated multijet contribution added (white histogram) are shown with the ($m_0$,$m_{1/2}$)=(100,150)~GeV signal point in the mSUGRA model with $\tan\beta=15$, $A_0=-2m_0$ and $\mu<0$ on top (dashed histogram). The multijet prediction is shown with $\pm 1$ standard deviation (s.d.) error bands (light brown area). The uncertainties in the prediction of the other backgrounds are not shown explicitely.
\label{fig:mettausel}}
\end{center}
\end{figure}

\begin{figure}[htbp]
\begin{center}
\includegraphics[scale=0.32,angle=00]{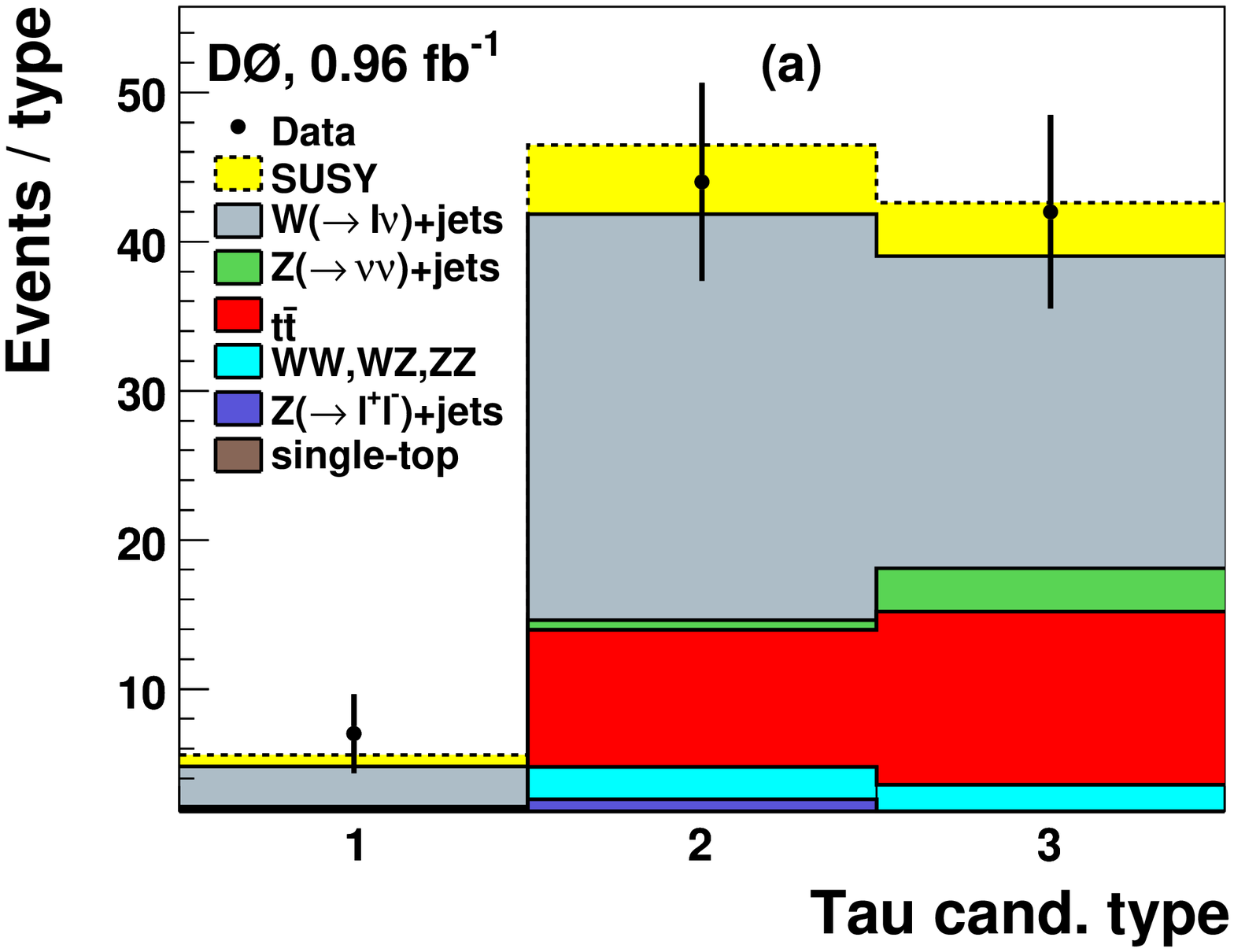}
\includegraphics[scale=0.32,angle=00]{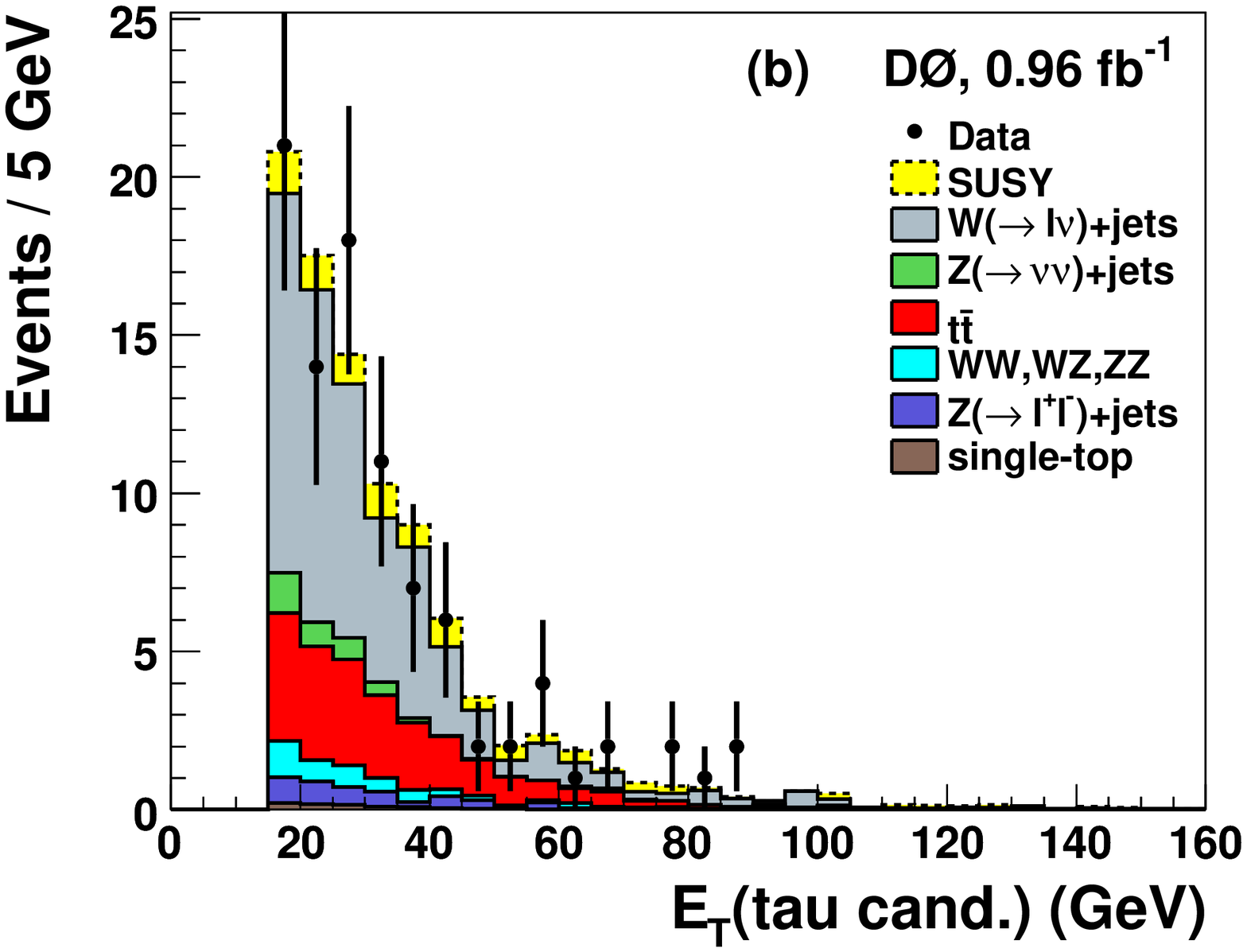}
\includegraphics[scale=0.32,angle=00]{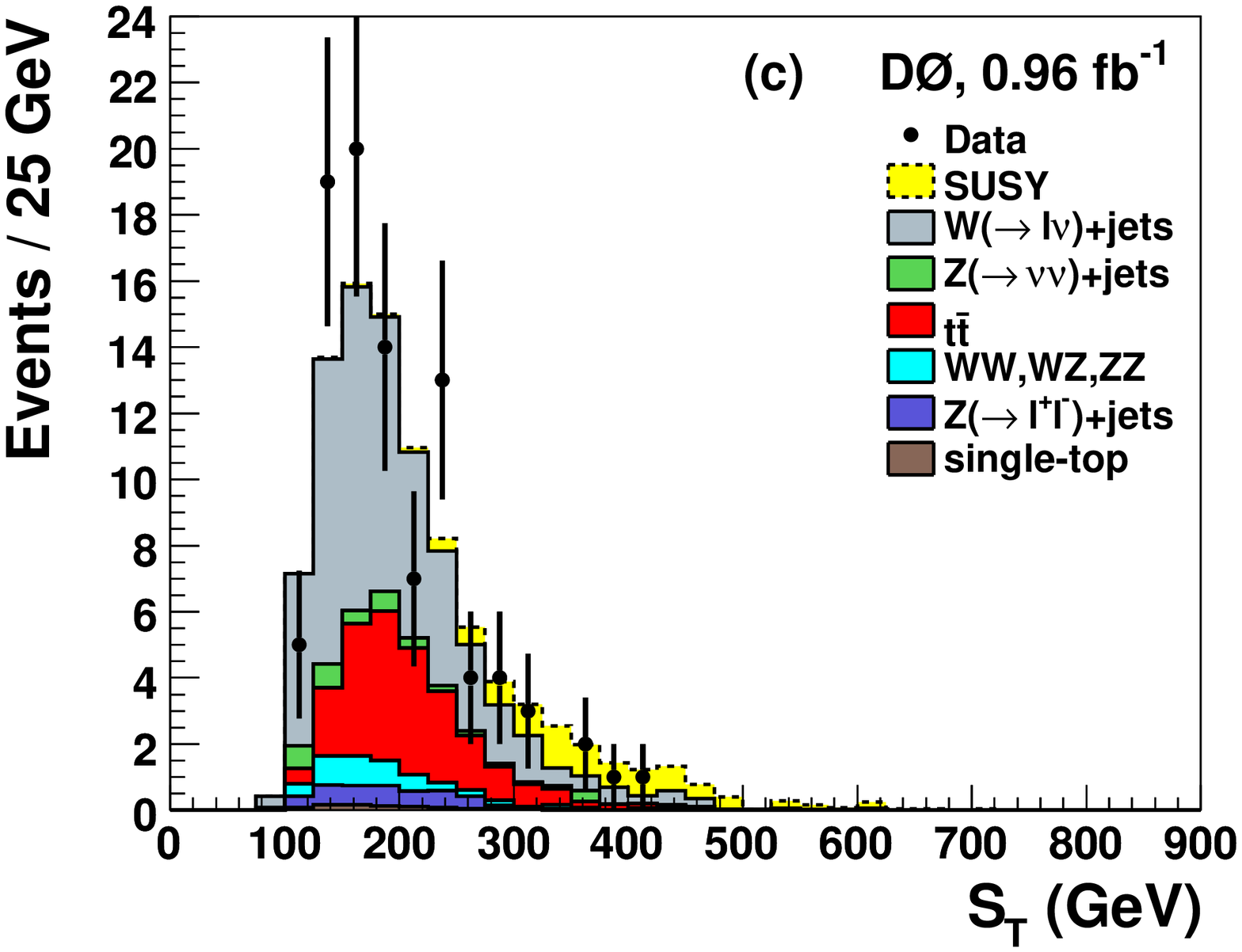}
\includegraphics[scale=0.32,angle=00]{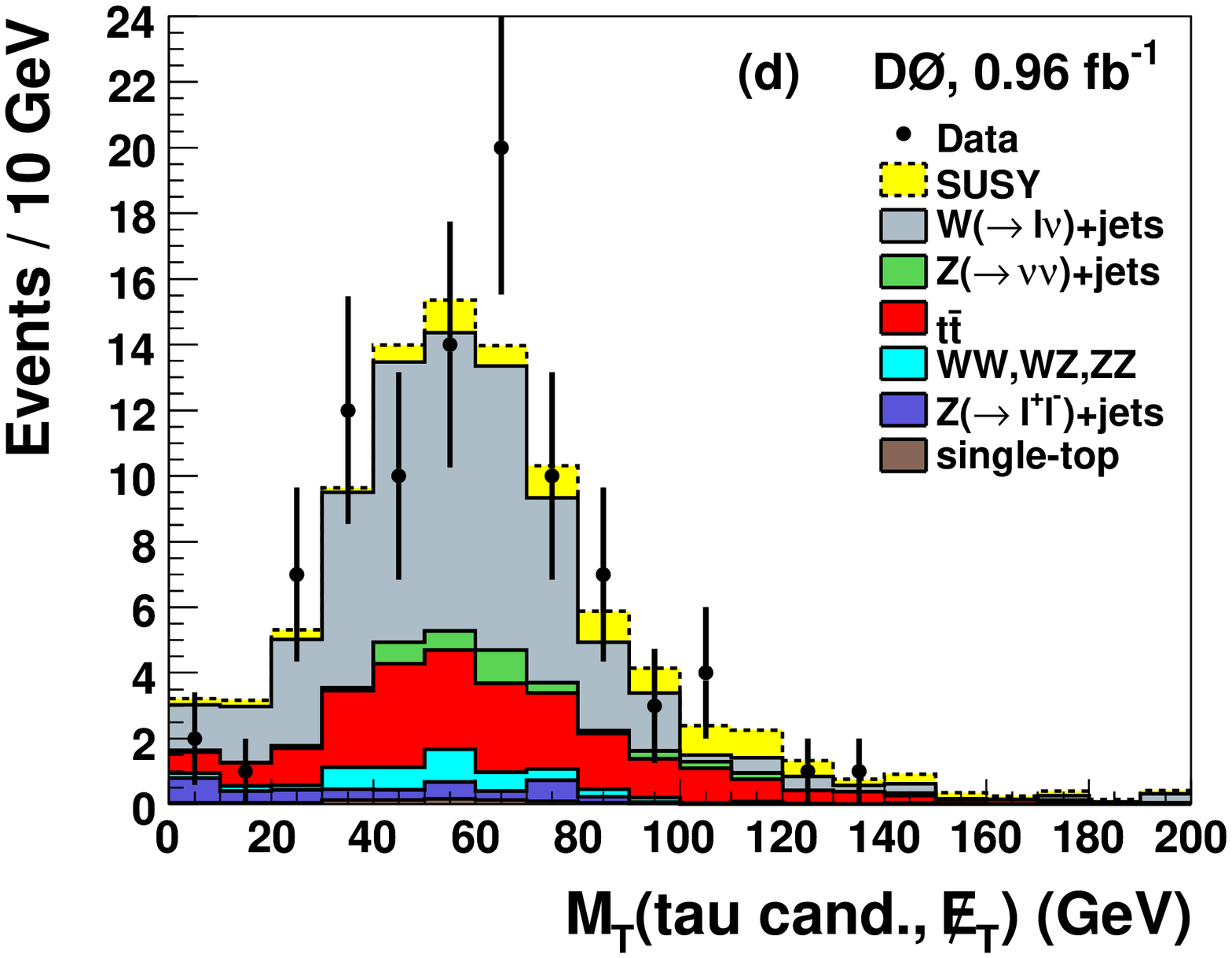}
\caption{Distributions of number of tau candidates per tau type (a), $E_T$ of tau candidates (b), $S_T$ (c)  and $M_T$($\tau_{cand}$,\met) (d) for events selected in the ``tau-dijet'' analysis after the tau selection is applied. Data (points with error bars) and SM predictions (plain histograms) without the estimated multijet contribution are displayed with the ($m_0$,$m_{1/2}$)=(100,150)~GeV signal point in the mSUGRA model with $\tan\beta=15$, $A_0=-2m_0$ and $\mu<0$ on top (dashed histogram).
\label{fig:tauseldijet}}
\end{center}
\end{figure}

At this stage of the analysis, events from $W$ and top quark pair production account for 85\% of the predicted background. According to the simulation, more than 50\% of tau candidates are due to a true hadronic tau at the particle level in all backgrounds.  Non-simulated multijet background is also sizable in the ``tau-multijet'' analysis. It enters the sample when one jet mimics the signal tau signature. It is mainly distributed at low \met\ as shown in Fig.\ \ref{fig:mettausel} for both analyses.
The background contribution from multijet events is estimated from data.
For this purpose, the tau identification criteria on $NN_j$ for type-1, type-2 and type-3 taus are dropped to define a loose superset of data. 
The multijet contribution to the signal sample is estimated from the observed event yields in the loose and in the signal sample and from the relative selection efficiency for events in the loose sample to also be part of this signal sample.
The selection efficiency is known separately for the signal-like events, which we simulate, and for multijet background, which we do not simulate but estimate from data. The latter efficiency is estimated from a data control sample enriched in multijet events. To this end, only the \met\ requirement is changed to select low \met\ events (\met\ $ \le 75$ GeV).
This technique is also known as the ``matrix method''.
In the ``tau-dijet'' analysis, after the \met\ cut at 75~GeV is applied, the number of events in each tau type is well predicted, as illustrated in Fig.\ \ref{fig:tauseldijet}.
In addition to normalisation, the shapes of quantity are also well described by the simulation. This includes observables used later on in the selection optimisation. For example, Fig.\ \ref{fig:tauseldijet} shows that the $E_T$ of the tau candidate is well described by the simulation, as well as $S_T$, a quantity indicative of the signal signature.
Angular correlations are also well described. For example, the transverse mass of the tau candidate and \met\ distributions are shown in Fig.\ \ref{fig:tauseldijet}. While the low edge of the distribution is shaped by kinematical cuts, it exhibits the Jacobian edge of the $W$ boson transverse mass at higher values.
This prevalence of tau candidates from $W$ decay in background events, coming either from top quark pair or $W+$jets production, is confirmed by the simulation.

Finally, the ``tau-dijet'' and the ``tau-multijet'' selections are combined with a logical {\texttt{OR}}, designated as the ``tau'' selection.  The signal selection is optimized with two additional cuts on \met\ and on $S_T$, as quoted in  Table \ref{tab:taucutflow}. The cut values are optimized by minimizing the expected upper limit on the cross section in the absence of signal computed with the modified frequentist $CL_s$ method \cite{cls:junk,cls:read}.

Statistical and systematic uncertainties are included in the $CL_s$ calculation. The systematic components are summarized in Table \ref{tab:tausyste}. The JES uncertainty dominates: it is 25\% for the background rate and 13\% for the signal rate. In comparison, uncertainties on jet resolution, on jet reconstruction and identification, and on the jet-vertex confirmation are negligible.
Uncertainties related to tau leptons come from three sources and are small. 
The tau energy correction systematic uncertainty comes from the  imperfect knowledge of the single pion response in the calorimeter in data. Measurements of the single pion response in data and their comparison with the simulation result in an uncertainty of  6\% in the single pion response used to derive the tau energy scale correction.
This translates into a 1.2\% uncertainty on the event yield.
The $NN_j$ learning method is limited by the statistics of the data sample used for background. The corresponding systematic uncertainty is therefore estimated by fluctuating the $NN_j$ input variables according to the original statistical uncertainty.
The uncertainties from the $NN$ against electrons are small in comparison.
The uncertainty on the tau reconstruction and identification efficiency is dominated by the track finding efficiency. It is determined to be 3.0\%.
The systematic uncertainty on the integrated luminosity measurement is 6.1\%~\cite{andeen} and the trigger efficiency uncertainty on the signal and background expectations is estimated to be 2\% \cite{verdier:2008}.
Based on {\sc mcfm}, a 15\% uncertainty is assigned to SM background cross sections.
The uncertainty on the acceptance due to the approximate knowledge of the PDF is computed by using the {\texttt{CTEQ6.1M}} PDF error sets.
The uncertainties in the modeling of the initial and final state radiation with {\sc pythia} are estimated by varying the scale $Q$ and the parton shower virtuality parameters \cite{verdier:2008}.

\begin{table}[htbp]
\renewcommand{\arraystretch}{1.2}
\caption{Relative systematic uncertainties on SM background expectations and signal events for the optimized ``tau'' analysis.
\label{tab:tausyste}}
\begin{ruledtabular}
\begin{tabular}{lrr}
Source          &  Background (\%) & ~~~~~~Signal (\%)~~~~~~\\
\hline
luminosity             &         6.1  ~ ~~~~~  &       6.1 ~~~ ~~~~~ \\
trigger                &           2  ~ ~~~~~  &         2 ~~~ ~~~~~ \\
\hline
jet energy scale       &          25  ~ ~~~~~  &        13 ~~~ ~~~~~ \\
jet resolution         &         1.0  ~ ~~~~~  &       1.0 ~~~ ~~~~~  \\
jet identification     &           1  ~ ~~~~~  &         1 ~~~ ~~~~~ \\
jet-vertex confirmation&           2  ~ ~~~~~  &         2 ~~~ ~~~~~ \\
\hline
tau identification     &         3.0  ~ ~~~~~  &       3.0 ~~~ ~~~~~ \\
tau energy corrections &         1.2  ~ ~~~~~  &       1.2 ~~~ ~~~~~ \\
tau $NN$ selection     &         1.2  ~ ~~~~~  &       1.2 ~~~ ~~~~~ \\
\hline
cross section          &          15  ~ ~~~~~  &         - ~~~ ~~~~~ \\
PDF (acceptance)       &           6  ~ ~~~~~  &         6 ~~~  ~~~~~ \\
ISR/FSR                &           6  ~ ~~~~~  &         6 ~~~  ~~~~~ \\
\end{tabular}
\end{ruledtabular}
\end{table}

\begin{figure}[htbp]
\begin{center}
\includegraphics[scale=0.32,angle=00]{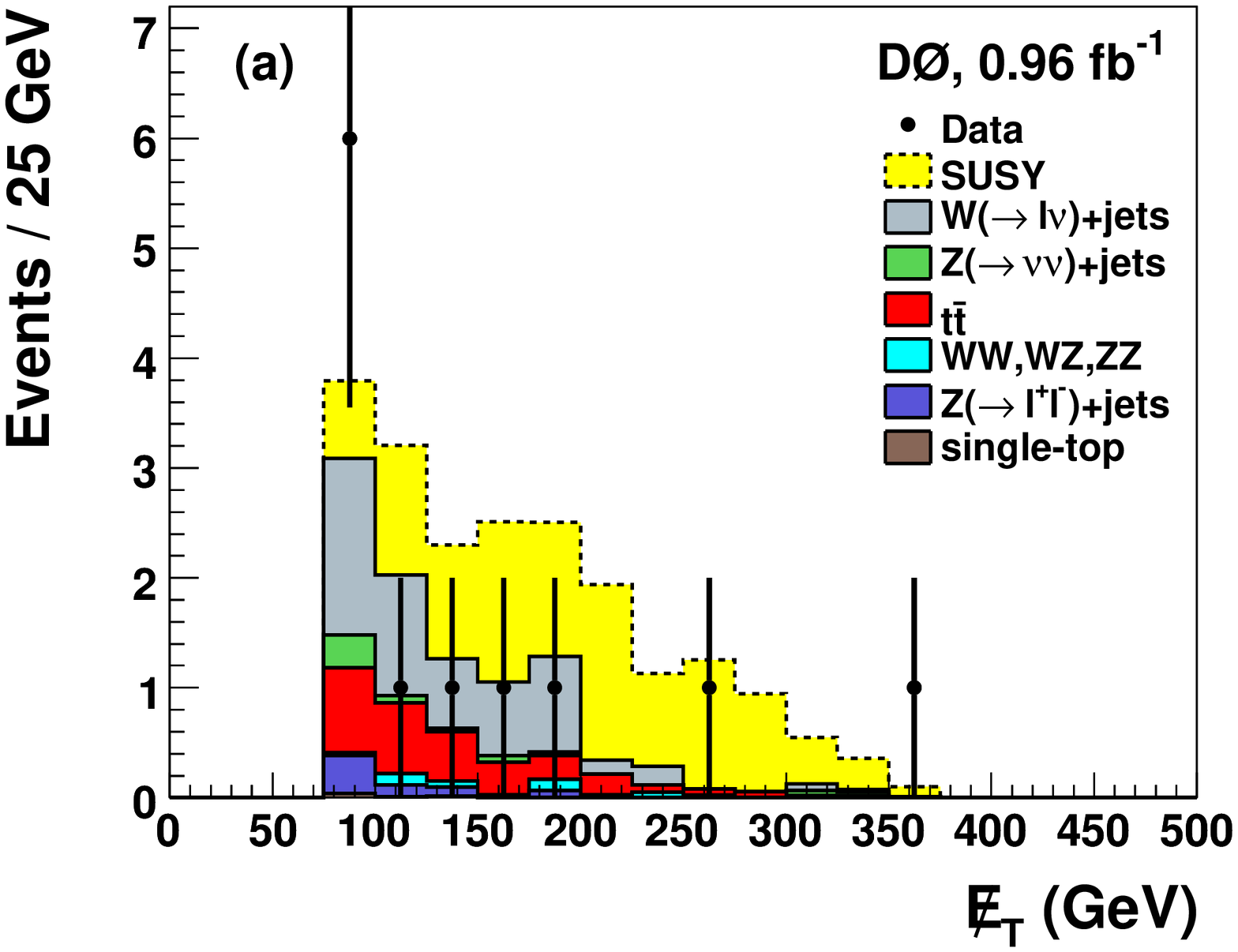}
\hfill
\includegraphics[scale=0.32,angle=00]{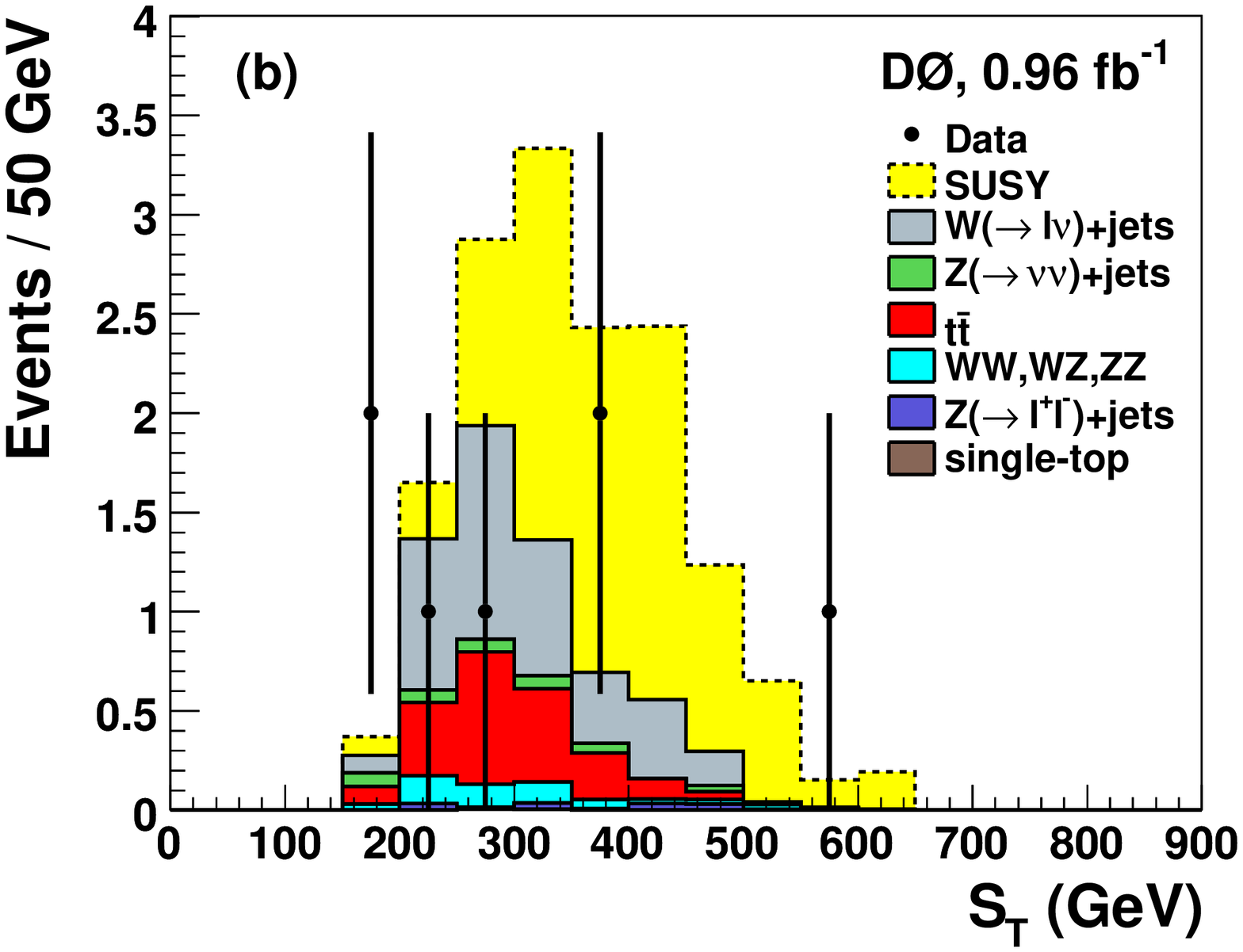}
\caption{Distributions of \met\ (a) and $S_T$ (b) with all the ``tau'' analysis requirements applied, except the one on the displayed quantity. Signal events for the point  ($m_0$, $m_{1/2}$)=(100, 150)~GeV in the mSUGRA model with $\tan\beta=15$, $A_0=-2m_0$ and $\mu<0$ are displayed on top of the background expectations. The multijet background is found to be small and it is therefore not shown explicitly.
\label{fig:marginal100150}}
\end{center}
\end{figure}

Three data events are selected by the ``tau'' analysis. This is in good agreement with the SM expectation of $2.3 \pm 0.4 {\rm (stat.)} \pm 0.7 {\rm (syst.)}$ events. Top quark pair production and $W(\to \ell \nu)+$jets events are the dominant background. 
The number of multijet events ($N_{\rm QCD}^{\rm sel}$) selected by the ``tau'' analysis is estimated from data with the matrix method. Statistical and systematic uncertainties are included in the calculation, leading to $N_{\rm QCD}^{\rm sel} = 0.1 \pm 0.6 $ events. 
The number of events expected for the signal point ($m_0$,$m_{1/2}$)=(100,150)~GeV is $6.5 \pm 0.6 {\rm(stat.)} \pm 1.1 {\rm(syst.)}$. 
In this case, tau candidates are predicted to arise from hadronic tau lepton decays 95\% of the time.
The distributions of $\met$ and $S_T$ are shown in Fig.\ \ref{fig:marginal100150} including the signal expectations assuming ($m_0$,$m_{1/2}$)=(100,150)~GeV. 
This point has a striking mass hierarchy where the $\tilde \tau _1^\pm$ is a few GeV lighter than the $\tilde \chi_2^0$. In this configuration, the lepton coming from the $\tilde \chi_2^0$ decay is mostly undetectable. Such mass configurations are difficult to detect in a search for direct $\tilde \chi_1^\pm \tilde \chi_2^0$ production in the three lepton final states \cite{pap:3leptons}. As shown below, this analysis is sensitive to this point at the 95\% C.L., whereas it is not excluded by the preliminary limits set on the slepton mass ($m_{\tilde e} > 100$~GeV, $m_{\tilde \mu} > 97$~GeV and  $m_{\tilde \tau} > 93$~GeV) and chargino mass ($m_{\tilde \chi^\pm} > 103$~GeV) by direct searches at LEP2~\cite{lep2limits}.

\begin{figure}[htbp]
\begin{center}
\includegraphics[scale=0.37,angle=00]{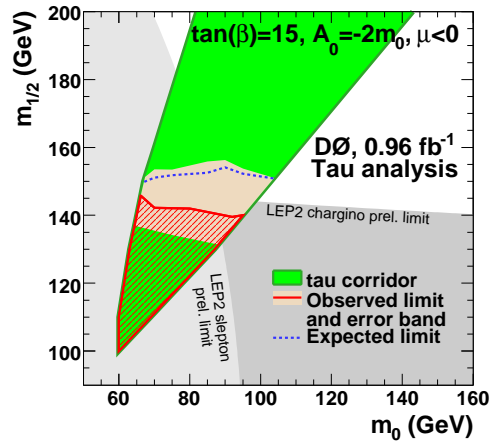}
\caption{In the \m0$-$\mf\ plane, expected and observed limits set by the ``tau'' analysis at the 95\% C.L. in the mSUGRA model assuming $\tan \beta = 15$, $A_0 =-2 m_0$ and $\mu <0$. Limits are derived in the ``tau corridor'' only, which is represented on top of the regions excluded by LEP2 slepton and chargino searches. 
The observed limit is shown with a variation of the signal cross section by $\pm 1$ s.d. 
The minimal and maximal limits are the boundaries of the lighter area within the darker area of the ``tau corridor''.
\label{fig:exclusionmzmf}}
\end{center}
\end{figure}

From the number of data events, the SM expectation and the signal yields, upper limits on the production cross section are derived at the 95\% C.L. The mSUGRA model with enhanced tau lepton final states is taken as a reference and only the ``tau corridor'' is explored.
Adding the estimated multijet event contribution to the SM background expectation does not change significantly the results. The multijet estimation  is therefore not taken into account.
Figure \ref{fig:exclusionmzmf} displays the expected and observed limits in the \m0$-$\mf\ plane in the mSUGRA model with $\tan\beta = 15$, $A_0 =-2 m_0$ and $\mu <0$. In the region of interest, the analysis sensitivity is kinematically limited by the squark masses.
Uncertainties on the signal cross section come from two major sources.
The uncertainty from the PDFs is computed by summing quadratically the forty individual {\texttt{CTEQ6.1M}} error contributions.
Uncertainties due to the imperfect knowledge of the renormalisation and factorisation scale are computed by varying by a factor two the default scale value $Q$ in {\sc prospino}. It is taken as  $Q= m_{\tilde q}, ~ (m_{\tilde q}+ m_{\tilde g})/2 ~{\rm and} ~ m_{\tilde g}$, where $m_{\tilde q}$ and $m_{\tilde g}$ are the masses of the squark and the gluino, respectively for $\tilde q \tilde q$, $\tilde q \tilde g$ and $\tilde g \tilde g$ production.
Renormalisation and factorisation scale uncertainties are typically 20\% or less and PDF uncertainties can be as high as 40\%.
They are added in quadrature. The result is translated directly into a minimal observed limit and a maximal observed limit. They represent a band on the observed limit, computed with the default PDF and $Q$ scale, as displayed in Fig.\ \ref{fig:exclusionmzmf}.
In the case of the nominal signal cross sections, the ``tau'' analysis excludes squark masses, averaged over eight squark species, up to 340~GeV. 
The expected limit reaches the indirect limits from LEP2 chargino and slepton searches.

The most recently published search for squarks and gluinos at D0 \cite{verdier:2008} in events with jets and large missing transverse energy using $2.1\;\rm fb^{-1}$ of data has not been specifically analyzed for tau lepton decays. However, hadronic tau lepton decays can be reconstructed as jets and we have studied the sensitivity of these analyses to tau lepton final states and combined them with the present ``tau'' analysis to improve sensitivity.

The published analyses are called ``dijet'', ``3-jets'' and ``gluinos'', each requiring an increasing number of jets. The ``gluino'' analysis requires four jets and it is not sensitive to the present signal topology. 
Regarding preselection and, jets and \met\ selection, the ``dijet'' and ``3-jets'' analyses are identical to the ``tau-dijet'' and ``tau-multijet'' analyses, respectively (see Table~\ref{tab:taucutflow}). 
The ``dijet'' and ``3-jets'' analyses veto high $p_T$ electrons and muons whereas the ``tau'' analysis rejects electrons and muons faking hadronic tau lepton decays. In the case of the ``dijet'' and ``3-jets'' analyses, the optimisation is performed on the pair of requirements $H_T$ and \met. The optimized selections are $H_T \ge 325$~GeV  and \met\ $\ge 225$~GeV for the ``dijet'' analysis and $H_T \ge 375$~GeV  and \met\ $\ge 175$~GeV for the ``3-jets'' analysis.
Overlaps between selections are taken into account by defining a logical \texttt{AND} between them.
While the ``tau'' analysis is limited to 0.96~\invfb, the ``dijet'' and ``3-jets'' analyses include an additional $1.17\;\rm fb^{-1}$ of data collected from June 2006 through August 2007. This leads us to consider seven exclusive channels within the ``early'' dataset (0.96~\invfb) and three exclusive channels in the ``late'' dataset ($1.17\;\rm fb^{-1}$). 
Table \ref{tab:10channeltau} details the ten channels of this combination and gives the number of selected and expected events for the SM backgrounds and for two signal points. 
Systematic uncertainties, as well as statistical uncertainties, are included in the $CL_s$ computation. 

\begin{table*}
\begin{center}
\begin{minipage}{\textwidth}
\renewcommand{\arraystretch}{1.2}
\caption{Exclusive channels used in the ``tau'', ``dijet'' and ``3-jets'' analysis combination in the mSUGRA model assuming $\tan\beta = 15$, $A_0 =-2 m_0$ and $\mu <0$. 
``Early'' and ``late'' refer to the two data taking periods used in this Letter.
The flag ``yes'' or ``no'' means events passing the corresponding analysis are accepted or rejected for this channel respectively. The number of selected data, the SM expectations and the number of signal events for the mSUGRA parameters ($m_0$,$m_{1/2}$)=(100,150)~GeV and ($m_0$,$m_{1/2}$)=(80,170)~GeV are also given. The first quoted uncertainty is statistical and the second one is systematic.
\label{tab:10channeltau}}
\begin{ruledtabular}
\begin{tabular}{rccccccccc}
{Chan.} & {early}& {early}& {early} & {late}& {late} & {Data} &{SM}&{Signal}&{Signal}\\ 
            & ``dijet'' & ``3-jets'' & ``tau'' & ``dijet'' & ``3-jets'' &&&(100,150)&(80,170)\\
\hline
~~1~~ &yes&no &no  & - & - & 4 & 4.2 $\pm$ 0.5 $\pm$ 1.1 & 6.7 $\pm$ 0.6 $\pm$ 1.0 & 3.5 $\pm$ 0.3 $\pm$ 0.6 \\
~~2~~ &no &yes&no  & - & - & 2 & 3.9 $\pm$ 0.4 $\pm$ 0.8 & 2.0 $\pm$ 0.4 $\pm$ 0.3 & 1.5 $\pm$ 0.2 $\pm$ 0.3 \\
~~3~~ &no &no &yes & - & - & 1 & 1.1 $\pm$ 0.3 $\pm$ 0.4 & 2.1 $\pm$ 0.4 $\pm$ 0.4 & 0.8 $\pm$ 0.2 $\pm$ 0.2 \\
~~4~~ &yes&yes&no  & - & - & 0 & 0.9 $\pm$ 0.2 $\pm$ 0.3 & 1.3 $\pm$ 0.3 $\pm$ 0.3 & 0.8 $\pm$ 0.2 $\pm$ 0.2 \\
~~5~~ &yes&no &yes & - & - & 0 & 0.1 $\pm$ 0.1 $\pm$ 0.1 & 1.1 $\pm$ 0.3 $\pm$ 0.2 & 0.4 $\pm$ 0.1 $\pm$ 0.1 \\
~~6~~ &no &yes&yes & - & - & 2 & 1.0 $\pm$ 0.3 $\pm$ 0.4 & 2.5 $\pm$ 0.4 $\pm$ 0.5 & 1.4 $\pm$ 0.2 $\pm$ 0.3 \\
~~7~~ &yes&yes&yes & - & - & 0 & 0.1 $\pm$ 0.1 $\pm$ 0.1 & 0.8 $\pm$ 0.2 $\pm$ 0.2 & 0.2 $\pm$ 0.1 $\pm$ 0.1 \\
~~8~~ & - & - & - &yes&no  & 4 & 5.1 $\pm$ 0.8 $\pm$ 1.3 & 9.1 $\pm$ 0.8 $\pm$ 1.4 & 4.5 $\pm$ 0.3 $\pm$ 0.7 \\
~~9~~ & - & - & - &no &yes & 2 & 4.1 $\pm$ 0.4 $\pm$ 0.8 & 5.3 $\pm$ 0.6 $\pm$ 0.8 & 3.3 $\pm$ 0.3 $\pm$ 0.5 \\
~10~~ & - & - & - &yes&yes & 3 & 0.8 $\pm$ 0.2 $\pm$ 0.3 & 2.4 $\pm$ 0.4 $\pm$ 0.5 & 1.2 $\pm$ 0.2 $\pm$ 0.3 \\
\end{tabular}
\end{ruledtabular}
\end{minipage}
\end{center}
\end{table*}

\begin{figure}[htbp]
\begin{center}
\includegraphics[scale=0.37,angle=00]{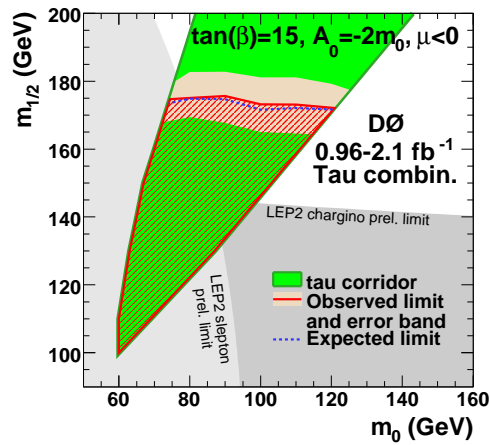}
\caption{In the \m0$-$\mf\ plane, expected and observed limits set by the combination  of the ``tau'', ``dijet'' and ``3-jets'' analyses. Limits are set at the 95\% C.L. in the mSUGRA model with $\tan\beta = 15$, $A_0 =-2 m_0$ and $\mu <0$.  Limits are derived in the ``tau corridor'' only,
which is represented on top of the regions excluded by LEP2 slepton and chargino searches.
The observed limit is shown with a variation of the signal cross section by $\pm 1$ s.d. 
The minimal and maximal limits are the boundaries of the lighter area within the darker area of the ``tau corridor''.
\label{fig:exclu10chatau}}
\end{center}
\end{figure}

\begin{figure}[htbp]
\begin{center}
\includegraphics[scale=0.37,angle=00]{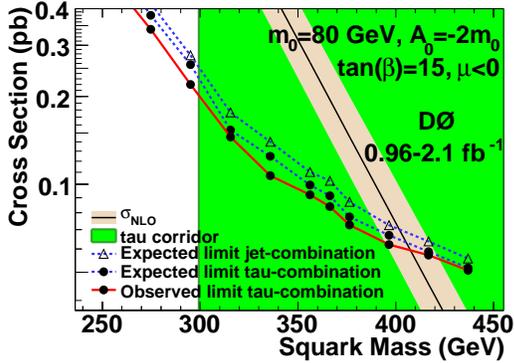}
\caption{Observed and expected 95\% C.L. upper limit on squark and gluino pair production cross sections with the combination of the ``tau'', ``dijet'' and ``3-jets'' analyses (labelled ``tau-combination'') for $m_0=80$~GeV in the mSUGRA model with $\tan\beta = 15$, $A_0 =-2 m_0$ and $\mu <0$. The expected cross section upper limit obtained with the ``dijet'' and ``3-jets'' analyses alone is also displayed (``jet-combination''). 
The NLO production cross section with a $\pm 1$ s.d. band, representing the PDF, renormalisation scale and factorisation
 scale uncertainties is superimposed.
\label{fig:xseclimit}}
\end{center}
\end{figure}

Final limits are explored in the ``tau corridor''  in the mSUGRA model with $\tan\beta = 15$, $A_0 =-2 m_0$ and $\mu <0$ as shown in Fig.\ \ref{fig:exclu10chatau}. The sensitivity of this combination reaches squark masses up to 408~GeV and the observed limit reaches 410~GeV for ($m_0$,$m_{1/2}$)=(90,176)~GeV or ($m_0$,$m_{1/2}$)=(110,173)~GeV. Although the ``tau'' analysis is performed on a dataset half the size of the ``dijet'' and ``3-jets'' analyses, 
Fig.\ \ref{fig:xseclimit} shows the relative gain in the production cross section upper limit is 10\% if the ``tau'' analysis is included in the combination.
This relative gain is 33\% if datasets of equal integrated luminosity are considered by each analysis.

The mass difference between the \neud\ (or the \chau), the \stu\ and the \neuu\ vary inside the ``tau corridor'', leading to different event kinematics.  
Model points close to the high $m_0$ border of the ``tau corridor'' predict that the \stu\ mass is a few GeV below the \neud\ mass, leaving the lepton produced in the \neud\ decay with a low $p_T$.
Going towards lower $m_0$, slepton masses decrease and the mass difference \neud\ $-$\ \stu\ increases, while the \stu$ - $\neuu\ mass difference decreases. Neighboring the LEP2 slepton limit, the point ($m_0$,$m_{1/2}$)=(80,170)~GeV exhibits similar mass differences between the \neud\ and the \stu\ (29 GeV), and between the \stu\ and the \neuu\ (27 GeV). This point, with a squark mass of 396~GeV and a production cross section of 0.08~pb, is excluded by the ``dijet'', ``3-jets'' and ``tau'' analysis combination. 
We have explored these extreme mass configurations, as well as intermediate ones, and we exclude them.

%%%%%%%%%%%%%%%%%%%%%%%%%%%%%%%%%%%%%%%%%%%%%%%%%%%%%%%%%%%%%%%%%%%%%%%%%%%%%%%
%\section{Conclusion}
%%%%%%%%%%%%%%%%%%%%%%%%%%%%%%%%%%%%%%%%%%%%%%%%%%%%%%%%%%%%%%%%%%%%%%%%%%%%%%%

In conclusion, squark pair production was searched for in events with jets, tau lepton(s) and
large missing transverse energy using $0.96$~fb$^{-1}$ of D0 data recorded at a
center of mass energy of 1.96\,TeV during the Run~II of the
Fermilab Tevatron Collider. A dedicated selection based on identified hadronic tau lepton decays 
has been developed. 
It was further combined with selections based only on jets and \met\ signatures
and analyzing a superset of data that is twice as large (2.1~\invfb).
No evidence for signal was observed in the combined analysis.
The result of this search has been interpreted in
terms of exclusion in the mSUGRA model with $\tan\beta=15$,
$A_0=-2m_0$ and $\mu<0$ enhancing final states with tau leptons.
The region of the parameter space where tau leptons are explicitly produced, the so-called ``tau corridor'', is investigated.
The highest excluded squark mass at 95\% C.L. is 410~GeV.

This search is the first to explore supersymmetric models in tau lepton final states in a multijet environment at the Fermilab Tevatron Collider. The combination of tau and inclusive analyses shows the highest sensitivity to the signature of interest. 
However, the ``tau'' analysis by itself is a crucial ingredient in order to 
obtain detailed insight into the nature of any signature from SUSY or other new physics that could soon be discovered at the Fermilab Tevatron Collider or at the CERN Large Hadron Collider.
It is also important to complement other searches such as direct
searches for chargino-neutralino production in the trilepton final state where one lepton could be kinematically undetectable.

%%%%%%%%%%%%%%%%%%%%%%%%%%%%%%%%%%%%%%%%%%%%%%%%%%%%%%%%%%%%%%%%%%%%%%%%%%%%%%%%%%%%%%%%%%%%%%%%%%%%%%%%
% acknowledgement_paragraph_r2.tex                         5/15/09
%
We thank the staffs at Fermilab and collaborating institutions, 
and acknowledge support from the 
DOE and NSF (USA);
CEA and CNRS/IN2P3 (France);
FASI, Rosatom and RFBR (Russia);
CNPq, FAPERJ, FAPESP and FUNDUNESP (Brazil);
DAE and DST (India);
Colciencias (Colombia);
CONACyT (Mexico);
KRF and KOSEF (Korea);
CONICET and UBACyT (Argentina);
FOM (The Netherlands);
STFC and the Royal Society (United Kingdom);
MSMT and GACR (Czech Republic);
CRC Program, CFI, NSERC and WestGrid Project (Canada);
BMBF and DFG (Germany);
SFI (Ireland);
The Swedish Research Council (Sweden);
CAS and CNSF (China);
and the
Alexander von Humboldt Foundation (Germany).
%
   % input acknowledgement

\end{document}